\documentclass[12pt]{article}
 \usepackage[utf8]{inputenc}
\usepackage[T1]{fontenc}
\usepackage{amsmath}
\usepackage{amsfonts}
\usepackage{dsfont}
\usepackage{caption}
\usepackage{subcaption}
\usepackage{longtable}
\usepackage{xcolor}
\usepackage{multirow}
\usepackage{float}
 \usepackage{amsmath}

\usepackage{graphicx}
\usepackage{color}

\usepackage[left=2.5cm,right=2.5cm,top=2.5cm,bottom=2.5cm]{geometry}
\begin{document}

\begin{center}
{\Large
	{\sc  Classification non supervisée des processus d'événements récurrents }
}
\bigskip

Génia Babykina$^{1}$, \underline{Vincent Vandewalle}$^{2}$
\bigskip

{\it
$^{1}$ ULR 2694 - METRICS - \'Evaluation des Technologies de Santé et des Pratiques Médicales, CHU Lille, University of Lille, F-59000 Lille, France ; evgeniya.babykina@univ-lille.fr \\
$^{2}$ Université Côte d’Azur, Inria, CNRS, LJAD, France ; vincent.vandewalle@univ-cotedazur.fr
}
\end{center}
\bigskip

{\bf R\'esum\'e.} Les événements de même type qui surviennent plusieurs fois pour un individu (événements récurrents) sont rencontrés dans différents domaines (défaillances des systèmes industriels, épisodes de chômage, conflits politiques, épisodes des maladies chroniques). L’analyse de ces événements doit tenir compte de la dynamique complète du processus de récurrence et non seulement du nombre d’événements observés. Des modèles statistiques pour l’analyse des événements récurrents sont développés dans le cadre probabiliste de processus de comptage. Un des modèles couramment utilisé est celui d’Andersen-Gill, une généralisation du modèle de durée de Cox, qui suppose que l’intensité de base d’un processus de récurrence dépend du temps et est ajusté sur les covariables observées. Pour un individu $i$ avec le vecteur de covariables $\pmb{X}_i$, cette intensité a une forme suivante :  $\lambda_{i}(t ; \pmb{\theta}) =\lambda_0(t)\exp\left(\pmb{X}_i\pmb{\beta}\right)$. L'intensité de base peut être écrite paramétriquement sous forme de Weibull $\lambda_0(t)=\gamma_{1}\gamma_{2} t^{\gamma_{2}-1}$, avec $\gamma_1$ paramètre d'échelle et $\gamma_2$ paramètre de forme. Néanmoins, les covariables observées sont souvent insuffisantes pour expliquer l’hétérogénéité observée dans les données. C’est notamment souvent le cas dans les études portant sur les patients dans le cadre d’études cliniques. Dans cet article un modèle de mélange pour l’analyse d’événements récurrents est proposé. Ce modèle permet de tenir compte de l’hétérogénéité inexpliquée et de réaliser la classification des individus vis-à-vis leur processus de récurrences. Au sein de chaque classe, l’intensité du processus de récurrences est spécifiée paramétriquement et dépend des variables explicatives.  Ainsi, l'intensité devient spécifique à la classe $k$  : $\lambda_{ik}(t ; \pmb{\theta}_k) =\gamma_{1k}\gamma_{2k} t^{\gamma_{2k}-1}\exp\left(\pmb{X}_i\pmb{\beta}_k\right)$. Les paramètres du modèle sont estimés par la méthode de Maximum de vraisemblance à l’aide de l’algorithme EM. Le critère BIC est utilisé pour choisir le nombre optimal des classes. La faisabilité du modèle est vérifiée par les simulations de Monte Carlo. Une application sur des données réelles portant sur les re-hospitalisations de personnes âgées est proposée. La faisabilité du modèle proposé est vérifiée empiriquement (l’algorithme d’optimisation converge, les estimations obtenues sont non-biaisées). L’application aux données réelles permet d’identifier deux classes de patients cliniquement cohérentes.

{\bf Mots-cl\'es.} \'Evénements récurrents, données censurées, modèle de mélange, Maximum de vraisemblance

\medskip

{\bf Abstract.} Event of the same type occurring several times for one individual (recurrent events) are present in various domains (industrial systems reliability, episodes of unemployment, political conflicts, chronic diseases episodes). Analysis of such kind of data should account for the whole recurrence process dynamics rather than only focusing on the number of observed events. Statistical models for recurrent events analysis are developed in the counting process probabilistic framework. One of the often-used models is the Andersen-Gill model, a generalization of the well-known Cox model for durations, which assumes that the baseline intensity of the recurrence process is time-dependent and is adjusted for covariates. For an individual $i$ with covariates $\pmb{X}_i$, the intensity is as follows: $\lambda_{i}(t ; \pmb{\theta}) =\lambda_0(t)\exp\left(\pmb{X}_i\pmb{\beta}\right)$. The baseline intensity can be specified parametrically, in a form of Weibull: $\lambda_0(t)=\gamma_{1}\gamma_{2} t^{\gamma_{2}-1}$, with $\gamma_1$ scale parameter et $\gamma_2$ shape parameter.  However, the observed covariates are often insufficient to explain the observed heterogeneity in data. This is often the case of clinical trials data containing information on patients. In this article a mixture model for recurrent events analysis is proposed. This model allows to account for unobserved heterogeneity and to cluster individuals according to their recurrence process. The intensity of the process is parametrically specified within each class and depend on observed covariates. Thus, the intensity becomes specific to class $k$: $\lambda_{ik}(t ; \pmb{\theta}_k) =\gamma_{1k}\gamma_{2k} t^{\gamma_{2k}-1}\exp\left(\pmb{X}_i\pmb{\beta}_k\right)$. The model parameters are estimated by the Maximum Likelihood method, using the EM algorithm. The BIC criterion is employed to choose the optimal number of classes. Model feasibility is verified by Monte Carlo simulations. An application to real data concerning hospital readmissions of elderly patients is proposed. The proposed model feasibility is empirically verified (the optimization algorithm converges, providing non-biased estimates). The real data application allows to identify two clinically relevant classes of patients. 

{\bf Keywords.} Recurrent events, censored data, mixture model, Maximum Liklelihood

\bigskip\bigskip

 
\section{Introduction}
Dans les applications cliniques, nous rencontrons souvent des événements récurrents : les rechutes de cancer, les ré-hospitalisations consécutives, les épisodes de déprime, \textit{etc.} Les études de ce type de données doivent tenir compte de la dynamique du processus et non seulement s'intéresser au nombre d'événements.  L'approche classique d'analyse statistique dans ces situations est le modèle  de Cox pour la durée et ses extensions aux plusieurs événements, telles que le modèle d'Andersen-Gill (Andersen  et Gill (1982)). Les covariables observées ne sont pas toujours suffisantes pour tenir compte de l'hétérogénéité individuelle observée. Dans ce contexte, la classification non supervisée permettrait d'identifier les partitions latentes des individus et d'expliquer cette latence par les caractéristiques facilement observables (socio-démographiques, cliniques, biologiques), permettant ainsi un suivi adapté des patients. Cette approche est possible via les modèles de mélange qui existent pour l'analyse de survie (temps jusqu'à un événement) [2], mais qui sont peu développés pour les événements récurrents. Nous proposons une approche qui combine le modèle paramétrique pour les événements récurrents et le cadre général des modèles de mélange. Les estimations sont réalisées par la méthode de maximum de vraisemblance avec l'algorithme EM d'optimisation. La faisabilité du modèle est évaluée par des simulations de Monte Carlo. Une application à des données réelles est réalisée.

\section{Méthodes }

\subsection{Modèle}
 En supposant la distribution de Weibull des instants entre les événements, pour un patient $i$ appartenant à une classe latente $k$, le processus de comptage d'événements  $\{N_i(t)\}_{t\geq 0}$ est caractérisé  par son intensité  $\lambda_{ik}(t)$ :
\begin{eqnarray}
\label{eq:intensity}
  \lambda_{ik}(t ; \pmb{\theta}_k) =\gamma_{1k}\gamma_{2k} t^{\gamma_{2k}-1}\exp\left(\pmb{X}_i\pmb{\beta}_k\right) ,
\end{eqnarray}
où  $\left(\gamma_{1k}, \gamma_{2k} \right)$ est le vecteur des paramètres de la distribution de Weibull et  $\pmb{\beta}_k$ est un vecteur des effets des covariables observées $\pmb{X}_i$ ; $\pmb{\theta}_{k}=\left(\gamma_{1k}, \gamma_{2k}, \pmb{\beta}_{k} \right)$.

Sous l'hypothèse de censure à droite non-informative, de présence de $K$ classes latentes avec $\pi_k$ : la probabilité d'appartenance à la classe $k$, la contribution à la vraisemblance d'un individu $i$ est une moyenne pondérée des vraisemblances spécifiques à la classe  : 
\begin{eqnarray*}
p(\pmb{t}_i|\pmb{X}_i ; \pmb{\theta}) 
& = & \sum_{k=1}^{K} \pi_k \prod_{j=1}^{n_i}\left[\lambda_{ik}\left(t_{ij}, \pmb{\theta}_k\right)\right] \times \exp \left(-\int_{0}^{\tau_i}\lambda_{ik}(u, \pmb{\theta}_k)du\right), \\
\nonumber
\end{eqnarray*}
où  $n_i$ est le nombre d'événements de l'individu $i$, survenus entre 0 et $\tau_i$ aux instants $\pmb{t}_i = \{t_{ij}\}_{j=1, \cdots, n_i}$.
Cette distribution est également appelée la distribution de mélange. 
\'Etant donné les instants d'événements  $\pmb{t}_i$ et les covariables $\pmb{X}_i$ pour un individu $i$, la probabilité \textit{a posteriori} d'appartenance à une classe peut être calculée à l'aide de formule de Bayes : 

\begin{eqnarray}
\label{eq:6}
\rho_{ik}=\mathds{P}(Z_{ik} = 1 | \pmb{t}_i,\pmb{X}_i ; \pmb{\theta}) = \frac{\pi_k p(\pmb{t}_i|\pmb{X}_i, Z_{ik} = 1 ; \pmb{\theta}_k)}{\sum_{\ell=1}^{K}\pi_{\ell}p(\pmb{t}_i|\pmb{X}_i, Z_{i\ell} = 1 ; \pmb{\theta}_{\ell})}.
\end{eqnarray}
Cette probabilité permet de classer l'individu, par exemple dans la classe dont cette probabilité est maximale. 

\subsection{Estimation de paramètres et sélection de modèles}

Les paramètres du modèle sont estimés par la méthode de Maximum de vraisemblance, en utilisant l'algorithme EM d'optimisation (Dempster \textit{et al.} (1977), Chauveau (1995)).
\`A l'étape $\mathrm{E}$ (\textit{expectation}) l'espérance de la vraisemblance complétée est calculée, elle nécessite la probabilité \textit{a posteriori} de l'Eq.(\ref{eq:6}) qui fournit l'espérance de données non-observées (classes latentes) conditionnellement aux observations pour les valeurs des paramètres à l'itération courante. \'A l'étape $\mathrm{M}$ (\textit{maximisation}) l'espérance de la vraisemblance complétée sachant les données observées et les paramètres courants est calculée. 
L'algorithme peut être initialisé par une partition aléatoire des individus en classes, permettant d'initier le vecteur de paramètres à estimer $\pmb{\theta}^{(0)}$. Les étapes $\mathrm{E}$ et $\mathrm{M}$ sont alternées jusqu'à la convergence.

\`A l'itération $h$ de l'algorithme $\mathrm{EM}$ l'espérance de la vraisemblance complétée est la suivante : 
\begin{eqnarray}
Q(\pmb{\theta} | \pmb{\theta}^{(h-1)}) &= &\sum_{i=1}^{n}\sum_{k=1}^{K} \rho_{ik}^{(h)} \log \left(\pi_k p\left(\pmb{t}_i|\pmb{x}_i, Z_{ik} = 1;\pmb{\theta}_k^{(h-1)}\right)\right) \\ \nonumber
&=& \sum_{k=1}^{K} Q\left(\pmb{\theta}_k | \pmb{\theta}^{(h-1)}\right),
\end{eqnarray}
avec $Q(\pmb{\theta}_k | \pmb{\theta}^{(h-1)}) := \sum_{i=1}^{n} \rho_{ik}^{(h)} \log (\pi_k p(\pmb{t}_i|\pmb{x}_i, Z_{ik} = 1;\pmb{\theta}_k))$.
\medskip

Les étapes $\mathrm{E}$ et $\mathrm{M}$ sont les suivantes : \\

\noindent  \textbf{\'Etape $\mathrm{E}$} : mise à jour des probabilités \textit{a posteriori} d'appartenance aux classes suivant l'Eq.(\ref{eq:6}) :
\begin{eqnarray*}
\rho_{ik}^{(h)}=\mathds{P}(Z_{ik} = 1 | \pmb{t}_i,\pmb{X}_i ; \pmb{\theta}^{(h-1)}) = \frac{\pi_k^{(h-1)} p(\pmb{t}_i|\pmb{X}_i, Z_{ik} = 1 ; \pmb{\theta}_k^{(h-1)})}{\sum_{\ell=1}^{K}\pi_{\ell}^{(h-1)}p(\pmb{t}_i|\pmb{X}_i, Z_{i\ell} = 1 ; \pmb{\theta}_{\ell}^{(h-1)})}.
\end{eqnarray*}

\noindent \textbf{\'Etape $\mathrm{M}$} : maximisation de $Q(\pmb{\theta} | \pmb{\theta}^{(r)})$ par rapport à  $\pmb{\theta}$. La formule de mise-à-jour de $\pi_k$ est $\pi_k^{(h)} = \frac{\sum_{i=1}^{n}\rho_{ik}^{(h)}}{n}$ ;  la mise-à-jour d'autres paramètres spécifiques aux classes peut être réalisée séparément : 
$$
\pmb{\theta}_k^{(h)} = \arg\max_{\pmb{\theta}_k} Q(\pmb{\theta}_k | \pmb{\theta}^{(h-1)}).
$$
Cette expression n'a pas de forme analytique, l'algorithme de Newton-Raphson est utilisé pour l'optimisation.

L'algorithme converge lorsque l'augmentation de la vraisemblance entre deux itérations successives est inférieure à un certain seuil. 

Le nombre de classes $K$ est choisi selon le critère BIC. Pour un modèle $\pmb{m}$ comportant $K$ classes and $d$ covariables, ce critère est défini comme : 
\begin{equation}
\mbox{BIC}(\pmb{m}) = \ell\left(\hat{\pmb{\theta}}^{\pmb{m}}\right) - \frac{K(3+d) -1}{2}\log n ,
\end{equation}
$\ell$ étant la log-vraisemblance, $\hat{\pmb{\theta}}^{\pmb{m}}$ l'estimateur du maximum de vraisemblance des paramètres du modèle $\pmb{m}$, et $K(3+d) - 1$  est le nombre de paramètres du modèle $\pmb{m}$. 


\subsection{Plan des simulations}

La faisabilité du modèle et l'algorithme d'estimations des paramètres ont été évalués via les simulations de Monté Carlo, tenant compte de la sensibilité des résultats à la taille d'échantillon ainsi qu'au  degré de séparation des classes. 
L'étude des simulations est réalisée pour deux classes latentes et deux variables explicatives. Les cas des classes bien séparées et des classes mélangées ont été considérés. La taille d'échantillon était fixée à 100 et 1000 individus. 
L'intensité de Eq.(\ref{eq:intensity}) était spécifiée sous forme de Weibull. La censure indépendante de Type I était considérée (les individus sont censurés à la fin de la période d'observation). La fin d'observation simulée est inspirée par les données réelles, correspondant à 1.99 sur l'échelle annuelle. Les variables explicatives considérées sont les suivantes : 
une issues de la distribution binomiale  $X_1 \sim \mathcal{B}\left(0.5\right)$ et l'autre issue de la distribution normale $X_2 \sim \mathcal{N}\left(0, 1\right)$.  

Pour le cas des \textit{classes bien séparées} les vrais paramètres de l'Eq.(\ref{eq:intensity}) sont les suivants : $\gamma_{11}=3$, $\gamma_{21}=2$, $\beta_{11}=0.4$, $\beta_{21}=-0.8$, $\gamma_{12}=1$, $\gamma_{22}=1$, $\beta_{12}=0.9$, $\beta_{22}=0.3$. Pour le cas des  \textit{classes mélangées}, les vrais paramètres sont les suivants : $\gamma_{11}=2$, $\gamma_{21}=2$, $\beta_{11}=0.5$, $\beta_{21}=-0.8$, $\gamma_{12}=1.5$, $\gamma_{22}=1.2$, $\beta_{12}=0.9$, $\beta_{22}=0.3$ . Les paramètres $\gamma_{1k}$ et $\gamma_{2k}$ correspondent respectivement au paramètre d'échelle (\textit{scale)} et le paramètre de forme (\textit{shape})  de la distribution de Weibull dans la classe $k$, $\beta_{1k}$ et $\beta_{2k}$ correspondent aux coefficients associées à $X_1$ et $X_2$ dans la classe $k$ respectivement.

\section{Résultats }
\subsection{\'Etude de simulations}

\begin{figure}[t]
     \centering
         \includegraphics[width=0.6\textwidth]{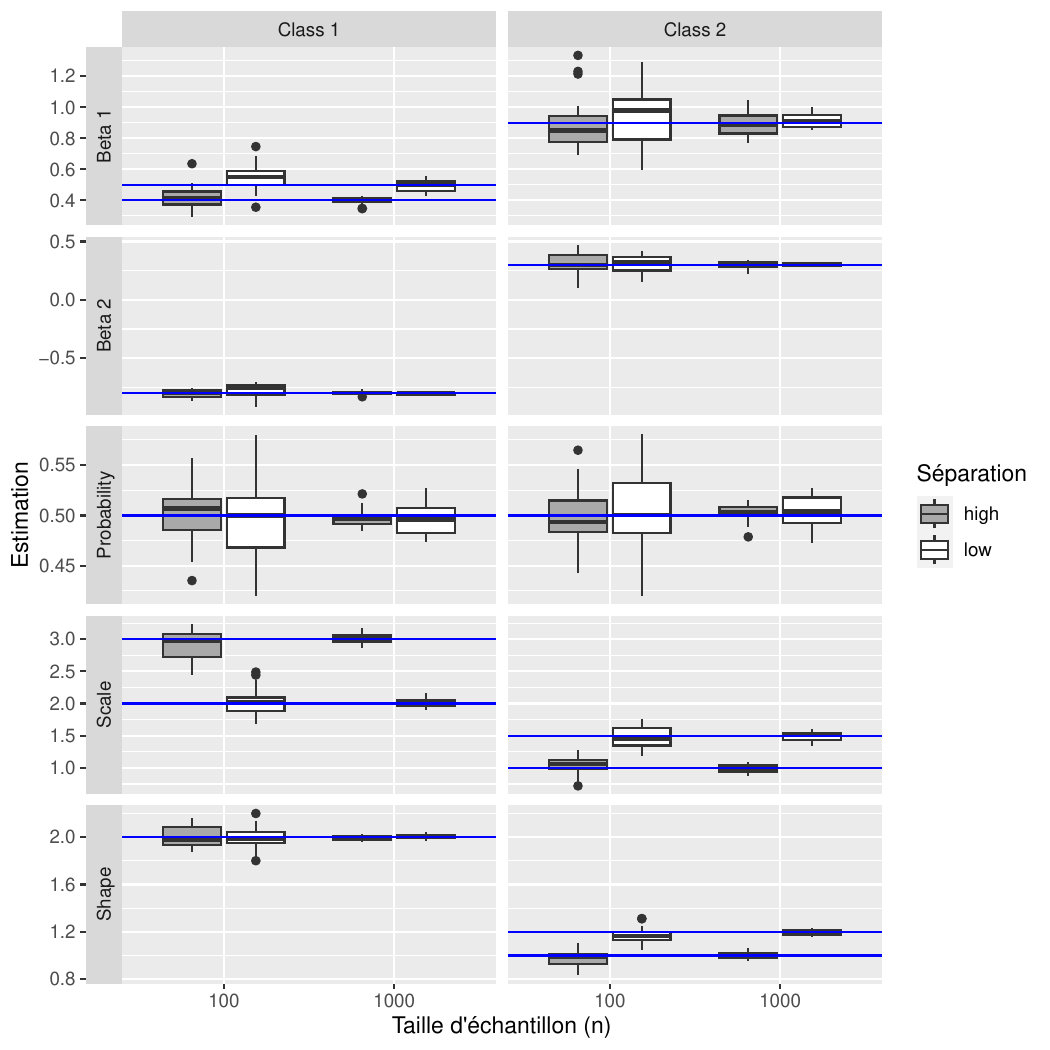}
           \caption{Données simulées : distributions des estimations de paramètres à partir de 20 simulations de Monte Carlo : les cas des classes bien séparées (\textit{high}) et des classes mélangées (\textit{low}) sont distingués, en bleu les valeurs des vrais paramètres pour les classes bien séparées et mélangées. }
            \label{fig:1}
\end{figure}

L'étude des simulations montre l'absence du biais dans les estimations des paramètres et la variance diminuant avec la taille d'échantillon qui augmente (voir Fig.\ref{fig:1}). L'erreur de classement est de l'ordre de 18\% pour les classes bien séparées et de 26\% pour les classes mélangées. 

\subsection{Application aux données réelles} 
Le modèle a été appliqué aux données réelles, la cohorte DAMAGE (Visade \textit{et al.} (2021)), utilisée pour l'évaluation  des facteurs de risque de décès et de re-hospitalisation après la sortie de l'hôpital des patients âgés. L'objectif de l'application est d'identifier les classes latentes des patients en termes de processus de ré-hospitalisation. 
Les données contiennent les caractéristiques cliniques  et socio-démographiques de 3088 patients âgés de plus de 75 ans, collectées pendant leur hospitalisation et après la sortie de l'hôpital : âge, genre, score de co-morbidité de Charlson, présence de démence, présence de cancer, lieu de résidence, mode de vie (isolé ou non), dates d'hospitalisations et de décès. 
Deux variables explicatives étaient inclues dans le modèle : âge et genre ; les autres variables disponibles étaient utilisées afin de caractériser les classes fournies par le modèle. 

Les estimations permettent d'identifier deux classes de patients en termes des hospitalisations consécutives (modèle optimal selon le critère BIC). Ces deux classes contiennent  2176 (70\%) et 912 (30\%) patients respectivement, les probabilités estimées d'appartenance aux classes sont respectivement  0.61 et 0.39. 
Les paramètres estimés sont fournis dans le Tab.\ref{tab:resEst2Class}, la description \textit{a posteriori} des classes figure dans le Tab.\ref{tab2}.

\begin{table}[ht]
\centering

\footnotesize
\begin{tabular}{r|rr}
  \hline 
 Paramètre & Classe 1 & Classe 2 \\ 
  \hline \hline
$\gamma_1$ & 0.52 & 2.38 \\ 
  $\gamma_2$ & 0.85 & 0.96 \\ 
  $\beta_{\text{Genre}}$ & -0.47 & -0.17 \\ 
   $\beta_{\text{\^Age}}$ & 0.28 & -0.08 \\ 
  $\pi$ & 0.61 & 0.39 \\ 
  $N$ & 2176 & 912 \\ 
  \% & 70.50 & 29.50 \\ 
   \hline
\end{tabular}
\caption{Application aux données réelles : estimation de paramètres du modèle à deux classes latentes : paramètre d'échelle (\textit{scale}) de  Weibull  $\gamma_1$, paramètre de forme (\textit{shape}) de Weibull $\gamma_2$, coefficients associés à genre et âge $\beta$ , probabilité marginale d'appartenance aux classes  $\pi$, taille des classes $N$ avec pourcentage.} 
\label{tab:resEst2Class}
\end{table}

\begin{table} 
\footnotesize{  
    \begin{tabular}{p{8cm}ccc}
      \hline  
     &      Classe  1         &       Classe 2         & p-value\\ 
 &      $N=2176$      &      $N=912$       &           \\ 
  
    \hline
    \hline     
   
    \`Age, N (\%): &                  &                  &   0.694  \\ 
$\qquad$(75-89) &   1493 (68.6\%)   &   633 (69.4\%)    &          \\ 
$\qquad$$>$89 &   683 (31.4\%)    &   279 (30.6\%)    &          \\ 
Genre, N (\%): &                  &                  &   0.005  \\ 
$\qquad$Homme &   707 (32.5\%)    &   345 (37.8\%)    &          \\ 
$\qquad$Femme &   1469 (67.5\%)   &   567 (62.2\%)    &          \\ 
Durée de suivi, Médiane [Q1;Q3] & 1.00 [1.00;1.00] & 0.77 [0.31;1.00] &  $<$0.001  \\ 
Nombre d'événements, Médiane [Q1;Q3] & 1.00 [1.00;2.00] & 3.00 [2.00;4.00] &   0.000  \\ 
Score de Charlson, N (\%): &                  &                  &  $<$0.001  \\ 
$\qquad$(0-2] &   983 (45.3\%)    &   310 (34.1\%)    &          \\ 
$\qquad$(2-5] &   1001 (46.2\%)   &   484 (53.2\%)    &          \\ 
$\qquad$$>$5 &   185 (8.53\%)    &   115 (12.7\%)    &          \\ 
Démence, N (\%) &   793 (36.5\%)    &   321 (35.2\%)    &   0.526  \\ 
Cancer, N (\%) &   291 (13.5\%)    &   169 (18.7\%)    &  $<$0.001  \\ 
Lieu , N (\%): &                  &                  &   0.326  \\ 
$\qquad$Maison &   1745 (80.3\%)   &   744 (81.8\%)    &          \\ 
$\qquad$Institution &   420 (19.3\%)    &   165 (18.1\%)    &          \\ 
$\qquad$Hôpital &    9 (0.41\%)     &    1 (0.11\%)     &          \\ 
Isolation, N (\%) &   178 (8.26\%)    &    84 (9.30\%)    &   0.387  \\ 
Décès, N (\%) &   495 (22.7\%)    &   522 (57.2\%)    &  $<$0.001  \\ 
Nombre d'événements, N (\%): &                  &                  &   0.000  \\ 
$\qquad$0 &   1557 (71.6\%)   &    0 (0.00\%)     &          \\ 
$\qquad$1 &   619 (28.4\%)    &   304 (33.3\%)    &          \\ 
$\qquad$2 &    0 (0.00\%)     &   376 (41.2\%)    &          \\ 
$\qquad$3 &    0 (0.00\%)     &   137 (15.0\%)    &          \\ 
$\qquad$4 + &    0 (0.00\%)     &    95 (10.4\%)    &           \\ 
 
    \hline

    \end{tabular}
          \caption{Application aux données réelles :  caractéristiques \textit{a posteriori} des classes latentes identifiées. Médianes et quartiles sont indiqués pour les variables quantitatives, effectifs et pourcentages sont indiqués pour les variables catégorielles. P-values des tests de comparaison des classes sont fournies (test de Mann-Whitney pour les variables quantitatives, test de chi-deux pour les variables catégorielles).}
      \label{tab2}}
    \end{table}

L'âge des individus n'est pas significativement différent entre les classes identifiées (les coefficients estimés sont 0.28 et -0.08 pour la classe 1 et 2 respectivement, la p-value du test de chi-deux est de 0.694) ; en revanche les classes sont différentes en termes de genre (il y a plus de femmes dans la classe 1, le coefficient estimé est -0.47 dans la classe 1 \textit{vs.} -0.17 dans la classe 2, la p-value du test de chi-deux est 0.005).

En ce qui concerne les variables non inclues dans le modèle, le score de co-morbidité de Charlson et la présence de cancer caractérisent les classes identifiées : la prévalence de cancer est de 13.5\% et de 18.7\% dans les classes 1 et 2 respectivement, les p-values des tests de chi-deux correspondants sont  $<0.001$ ; les patients de la classe 1 ont un score de Charlson moins élevé (moins de co-morbidités) que dans la classe 2 (8.53\% avec le score $>5$ \textit{vs.} 12.7\% respectivement, p-value du test de chi-deux $<0.001$). 
Dans la classe 1 la durée de suivi est significativement plus longue  (durée médiane de 1 an \textit{vs.} 0.77, \textit{i.e.} approximativement 9 mois dans la classe 2, la p-value du test de Mann-Whitney $<0.001$). Il y a moins de décès dans la classe 1  (22.7\% \textit{vs.} 57.2\% dans la  classe 2, la p-value du test de chi-deux $<0.001$), et les patients de la classe 1 ont moins d'hospitalisation répétées (100\% ont moins de 3 hospitalisations \textit{vs.} 25.4\% avec 3 hospitalisation ou plus dans la classe 2, p-value du test de chi-deux  $<0.001$).
Les paramètres de la distribution de Weibull estimés indiquent un processus de re-hospitalisations plus lent pour la classe 1 (les paramètres d'échelle et de forme correspondants sont 0.52 \textit{vs.} 0.85 et 2.38 \textit{vs.} 0.95 respectivement) 
\medskip

Pour résumer, les deux classes latentes identifiées indiquent \textbf{deux profils d'individus} : patients fragiles, avec un risque de décès plus élevé, plus souvent ré-hospitalisés, avec plus d'hommes et plus de co-morbidités, ainsi que la prévalence de cancer plus élevée (classe 2) et les patients plus robustes, avec le risque de décès moins élevé, plus de femmes, moins de co-morbidités, prévalence de cancer moins élevée et qui sont re-hospitalisés moins souvent (classe 1).

\section{Conclusion et perspectives}
Dans cet article, un modèle de classification dans le cadre d'analyse d'événement récurrents est proposé. Cette approche est bâtie sur un modèle paramétrique pour les événements récurrents, basé sur le processus de comptage, et sur un modèle de mélange. L'algorithme EM est proposé pour l'estimation des paramètres. La faisabilité du modèle est évaluée via les simulations de Monte Carlo. Les estimations fournies sont non-biaisées et ont une propriété de consistance constatée empiriquement. La performance du modèle est plus élevée lorsque les classes sont bien séparées. 
Une application aux données réelles porte sur la modélisation des hospitalisations des personnes âgées. Le modèle proposé permet d'identifier les classes des patents homogènes vis-à-vis le processus de re-hospitalisations. Ces classes sont cliniquement interprétables par les variables mesurées en début de suivi (co-morbidités, âge, genre) et par les variables non-mesurables au début de suivi (décès). 
En perspective, il serait intéressant de développer un modèle plus flexible permettant de tenir compte de différents distributions pour le processus de récurrences (Exponentiel, Gamma) ainsi que d'autres modèles pour les événements récurrents. 

\section*{Bibliographie }
\noindent Andersen, P. K., Gill, R. D. (1982), Cox's regression model for counting processes : a large sample study, {\it The Annals of Statistic},  pp. 1100--1120.\\

\noindent Dempster, A. P., Laird, N. M., et Rubin, D. B. (1977), Maximum likelihood from incomplete data via the EM algorithm, {\it Journal of the Royal Statistical Society: Series B (methodological)}, 39(1), pp. 1--22. \\

\noindent  Chauveau, D., (1995), A stochastic EM algorithm for mixtures with censored data, {\it Journal of Statistical Planning and Inference},  46(1), pp. 1–25.\\

\noindent Visade, F., Babykina, G., Puisieux, F., Bloch, F., Charpentier, A., Delecluse, C., ... et Beuscart, J. B. (2021), Risk factors for hospital readmission and death after discharge of older adults from acute geriatric units: taking the rank of admission into account, {\it Clinical Interventions in Aging}, pp. 1931-1941. \\

\end{document}